\documentclass[runningheads]{llncs}
\usepackage{graphicx}
\usepackage[misc]{ifsym} %used for the envelope
\usepackage{subfigure}
\usepackage{amssymb}
\usepackage{amsmath}
\begin{document}
\title{Calibrating Wayfinding Decisions in Pedestrian Simulation Models: The Entropy Map}
\titlerunning{Calibrating wayfinding in pedestrian simulation models}
\author{Luca Crociani\inst{1}\textsuperscript{(\Letter)}, Giuseppe Vizzari\inst{1} and Stefania Bandini\inst{1,2}}

\authorrunning{L. Crociani et al.}
\institute{\textsuperscript{1}Complex Systems and Artificial Intelligence research center\\ University of Milano-Bicocca, Viale Sarca, 336 - U14, 20126, Milan, Italy\\
\email{\{name.surname\}@unimib.it}\\
\textsuperscript{2}Research Center for Advanced Science and Technology\\ The University of Tokyo, 4-6-1, Komaba, Meguro-Ku, Tokyo, 153-8904, Japan\\
\email{}}

\maketitle              % typeset the header of the contribution
\begin{abstract}

This paper presents entropy maps, an approach to describing and visualising uncertainty among alternative potential movement intentions in pedestrian simulation models. In particular, entropy maps show the instantaneous level of randomness in decisions of a pedestrian agent situated in a specific point of the simulated environment with an heatmap approach. Experimental results highlighting the relevance of this tool supporting modelers are provided and discussed.
 
\end{abstract}

\section{Introduction \& Related Works}

Computer simulation of complex systems often employs stochastic models: implied randomness is a way to account for aspects that are potentially relevant to the overall phenomenon but cannot be explicitly considered to keep the model and the modelling phase manageable~\cite{BattyEditorial}.

Pedestrian and crowd behaviour simulation, for instance, requires considering different kinds of decisions, taken at distinct levels of abstraction, employing heterogeneous information and knowledge about the environment, from path planning~\cite{Crociani_IA2016} to the regulation of distance from other pedestrians and obstacles present in the environment\cite{CrocianiJCA,Crociani_SIMPAT2018}.

Exploring implications of randomness and situations of indecision, irresolution in case of choice among alternative lines of behaviour such as the exits from an environment in an emergency situation~\cite{LOVREGLIO201659}, can be a very significant step, with important implications of overall simulation results. 

This paper presents an approach to describing and visualising uncertainty among alternative potential movement intentions in pedestrian simulation models. As in the framework of probability theory~\cite{robinson2008entropy}, we use the concept of \textit{entropy} to provide a measure of uncertainty over the simulated space

The paper, first of all, describes a general decision making model for supporting wayfinding, which comes from previous work by the authors~\cite{Crociani_SIMPAT2018,Crociani_IA2016}. Then it introduces the notion of entropy map, as a practical tool to show the instantaneous level of uncertainty from the perspective of a pedestrian agent situated in a specific point of the simulated environment with an heatmap approach. Experimental results showing the potential usage of this practical tool supporting modelers will then be given and discussed. Conclusions and future developments end the paper.

\section{A Model for Plausible Wayfinding of, Pedestrians}
The multi-agent model designed for the simulation of pedestrian movement and route choice behaviour is thoroughly described in its components in~\cite{Crociani_SIMPAT2018,Crociani_IA2016}. The model of agent is composed of two elements, respectively devoted to the low level reproduction of the movement towards a target (i.e. the operational level, considering a three level model described in~\cite{Michon1985}) and to the decision making activities related to the next destination to be pursued (i.e. the route choice at the tactical level). Here we will report technical details only regarding the wayfinding component of the behavioural model, which are necessary to understand the analysis proposed in this paper. Moreover, for a proper understanding of these approaches, a brief description on the representation of the environment with different levels of abstractions will be also included. 

\subsection{The Representation of the Environment and the Knowledge of Agents}
The environment~\cite{WeynsDefJaamas} is discrete and modelled with a rectangular grid of 40 cm sided square cells, as usually applied in this context. The simulation scenario is drawn by means of several \emph{markers}. Basic markers of the scenario are \emph{start areas}, \emph{obstacles} and \emph{final destinations} --ultimate targets of pedestrians. To allow the wayfinding, two other markers are introduced. \emph{Openings} are sets of cells that divide the environment into regions, together with obstacles. These objects constitutes decision elements for the route choice, which will be denoted as \emph{intermediate targets}. Finally, \emph{regions} are markers that describe the type of the region: with them it is possible to design particular classes of regions (e.g. stairs, ramps).

This model uses the \emph{floor fields} approach~\cite{Burstedde2001}, spreading potentials from cells of obstacles and destinations to provide information about distances. The two types of floor fields are denoted as \emph{path field}, spread from each target, and \emph{obstacle field}, a unique field spread from all obstacle cells. In addition, a \emph{dynamic} floor field that has been denoted as \emph{proxemic field} is used to reproduce plausible distance towards the agents in low density situations.

The presence of intermediate targets allows the computation of a graph-like representation of the walkable space, based on the concept of \emph{cognitive map}~\cite{Tolman1948}. The algorithm for the computation of this data structure is defined in~\cite{crociani2014hybrid} and it uses the information of the floor fields associated to openings and final destinations. Recent approaches explores also the modelling of partial and incremental knowledge of the environment by agents (e.g.~\cite{Andresen2016}), but this aspect goes beyond the scope of the current work. The data structure identifies \emph{regions} (e.g. a room) as nodes of the labelled graph and \emph{openings} as edges. Overall the cognitive map allows the agents to identify their position in the environment and it constitutes a basis for the generation of an additional knowledge base, which will enable the reasoning for the route calculation. 

This additional data structure is denoted as \emph{Paths Tree} and it contains the free flow travel times\footnote{Calculated with their \emph{desired speed} of walking.} related to \emph{plausible} paths towards a final destination, from each intermediate target in the environment. The concept of plausibility of a path is encoded in the algorithm for the computation of the tree, which is thoroughly discussed in~\cite{DBLP:journals/ia/CrocianiPVB15}. 

Formally, for the choice of paths the agents access the information of a Paths Tree, generated from a final destination $End$, with the function $Paths(R, End)$. Given the region $R$ of the agent, this outputs a set of couples $\{(P_i,tt_i)\}$. $P_i = \left\lbrace \Omega_k, \ldots, End \right\rbrace$ is the ordered set describing paths starting from $\Omega_k$, belonging to $Openings(R)$ and leading to $End$. $tt_i$ is the respective free flow travel time.

\subsection{The Route Choice Model of Agents}

The objective of this component of the model is to enable agents to choose their path considering distances as well as the evolution of the dynamics. At the same time, the model must provide a sufficient variability of the results (i.e. of the paths choices) and a calibration over possible empirical data.

\begin{figure}[t]
\begin{center}
\includegraphics[width=.85\textwidth]{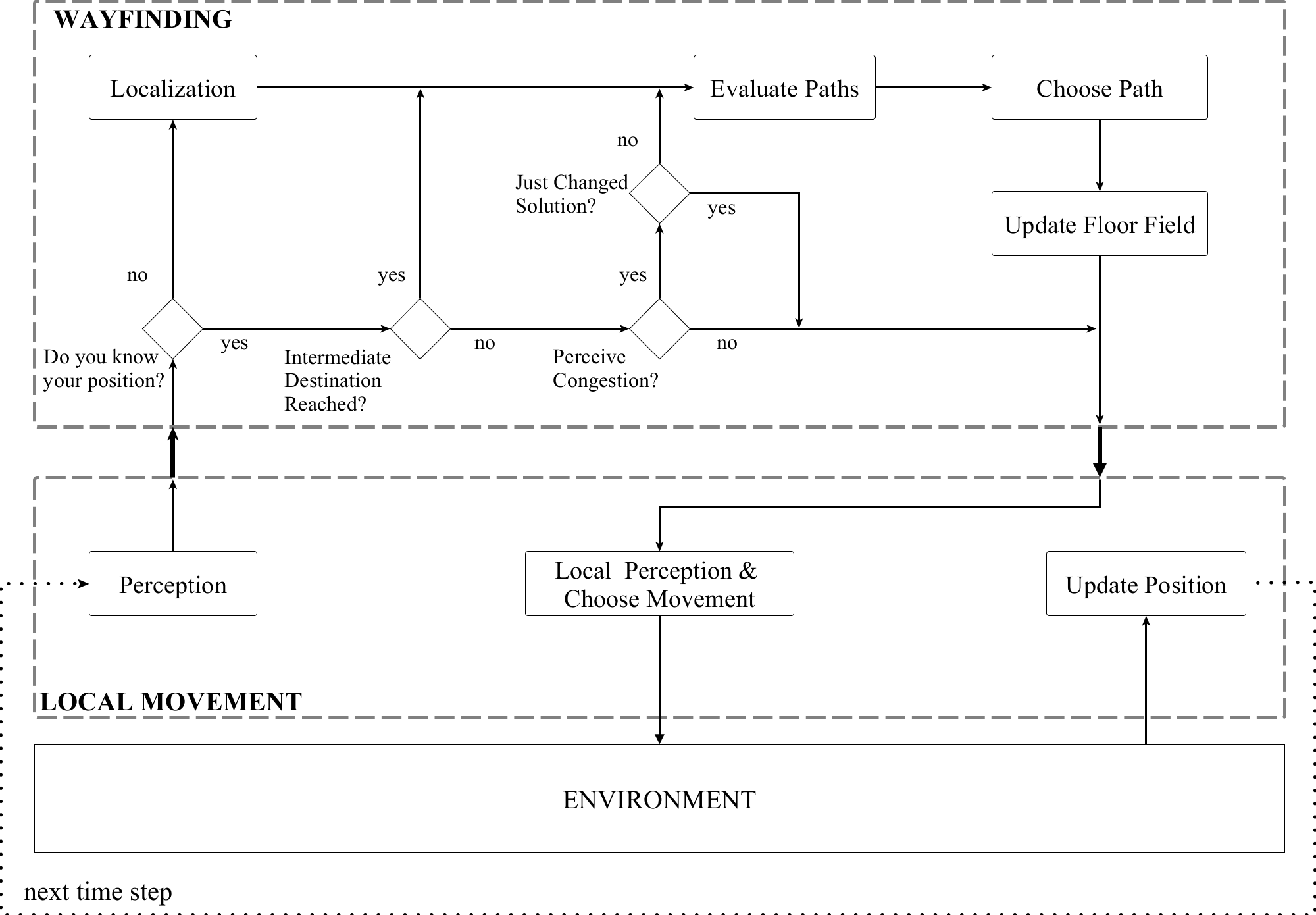}
\caption{The life-cycle of the agent, schematized among the two component of the agent architecture.}\label{fig:agent_lifeCycle}
\end{center}
\end{figure}

To understand the mechanisms designed in the model, the discussion must start with an overview of the agent life-cycle, illustrating which activity is performed and in which order. The workflow of the agent, encompassing the activities at operational and tactical level of behaviour at each time-step, is illustrated in Figure~\ref{fig:agent_lifeCycle}.

First of all, the agent performs a perception of its situation considering its knowledge of the environment, aimed at understanding its position and the markers perceivable from its region (e.g.\ intermediate targets). At the very beginning of its life, the agent does not have any information about the location, thus the first assignment to execute is \emph{localization}. This task analyses the values of floor fields in its physical position and infers the location in the Cognitive Map. Once the agent knows the region where it is situated, it loads the Paths Tree and evaluates possible paths towards its final destination. 

The evaluation has been designed with the concept of \emph{path utility}, assigned to each path to successively compute a probability to be chosen by the agent. The probabilistic choice of the path outputs a new intermediate target of the agent, used to update the reference to the floor field followed at the operational layer with the local movement. 

The scheme points out that the evaluation of the plan is not only performed at the beginning of the simulation. The wayfinding component is in fact activated at the beginning of every step, yet the plan is reconsidered only in two cases: (i) the agent has just entered into a new region or (ii) the current way that the agent is following has been perceived as congested. 

The first case means that the agent enters in a new region and it is able to perceive new information about its current path and the possible alternatives, thus it performs a new evaluation. In the second case, however, the evaluation is performed while reaching the intermediate target, if the path passing through this has been perceived congested. Basically this implies that until this congestion is perceived, the agents reconsider their decisions in favour of alternatives, thus assuming an uncertain behaviour in case of having multiple congested ways. To improve the dynamics in this situations, an \emph{inertia} mechanisms preserving the current decision is introduced. This is managed by means of two parameters, configuring two time intervals: a short one ($\tau_{short}$), to use right after a new decision has been taken, and a longer one ($\tau_{long}$), to be used if the evaluation led again to the current or previously chosen choice.

In particular, when an agent reconsiders the chosen plan due to congestion and then chooses again the original plan, then the choice will not be reconsidered before a longer period of time $\tau_{long}$. The rationale behind this modelling choice is that the agent has tried an alternative path, but eventually it turned out to be less convenient than initially expected. Therefore the agent will have some inertia and it will try to stand by the original plan at least for $\tau_{long}$, limiting erratic forth and back movements.

The utility-based approach fits well with the needs to easily calibrate the model and to achieve a sufficient variability of the results. %On the other hand, a structural difference that must be emphasized is that at the locomotion layer the reasoning is performed by means of the low level perception of the local situation (i.e. : position of targets, obstacles, local densities and eventual group members), composing a pure \emph{reactive} behaviour of the agent. Differently, the utility-based reasoning for the route choice is performed on the basis of more complex information structures and considering a more global perspective of the scene. Hence, the model at the tactical layer configures a \emph{cognitive} behaviour. 
The core functions of the wayfinding model are \emph{Evaluate Paths} and \emph{Choose Paths}, which will be now discussed.

%PATHS EVALUATION
\subsubsection{The Utility and Choice of Paths}
The function that computes the probability of choosing a path is exponential with respect to the utility value associated to it. This is completely analogous to the choice of movement at the operational layer:

\begin{equation}
Prob(P) = N \cdot e^{U(P)}
\end{equation}

The usage of the exponential function for the computation of the probability of choosing a path $P$ is a good solution to emphasize the differences in the perceived utility values of paths, limiting the choice of relatively bad solutions (that in this case would lead the agent to employ relatively long paths). $U(P)$ comprises the three observed components influencing the route choice decision, which are aggregated with a weighted sum:

\begin{small}\begin{equation}
U(P) = \kappa_{tt} Eval_{tt}(P) - \kappa_q Eval_{q}(P) + \kappa_f Eval_{f}(P)
\end{equation}\end{small}

where the first element evaluates the expected travel times; the second considers the \emph{queueing} (crowding) conditions through the considered path and the last one introduces a positive influence of perceived choices of nearby agents to pursue the associated path $P$ (i.e. imitation of emerging leaders). All the three functions provide values normalized within the range $[0,1]$, thus the value of $U(P)$ is included in the range $[-\kappa_q,\kappa_{tt}+\kappa_{f}]$.

In theory, there is no best way to define these three components: the usage of very simple functions as well as complicated ones might provide the same quality to the model. The only way to evaluate the reliability of this model, in fact, is with a validation procedure over some empirical knowledge. Hence, these three mechanisms have been designed with the main objective to allow the calibration over empirical datasets, preferring the usage of simple functions where possible.

\subsubsection{The Evaluation of Travelling Times}
To evaluate the travelling times relative to a path, the information about the travel time $tt_i$ of a path $P_i$ is firstly derived from the Paths Tree with $Paths(R, End)$ (where 
$End$ is the agent's final destination, used to select the appropriate Paths Tree, and $R$ is the region in which the agent is situated and it is used to select the relevant path $P_i$ in the Paths Tree structure) and it is integrated with the free flow travel time to reach the first opening $\Omega_k$ described by each path:

\begin{equation}
\mathit{TravelTime}(P_i) = tt_i + \frac{\mathit{PF}_{\Omega_k}(x,y)}{\mathit{Speed}_d}
\end{equation}

where $\mathit{PF}_{\Omega_k}(x,y)$ is the value of the path field associated to $\Omega_k$ in the position $(x,y)$ of the agent and $\mathit{Speed}_d$ is the \emph{desired velocity} of the agent, that can be an arbitrary value $\in \mathbb{R}$ (see~\cite{CrocianiJCA} for more details of this aspect of the model). 
 
The proposed function for the evaluation computes the minimum value of travelling times over the possible paths as the following:

 \begin{equation}
 \mathit{Eval}_{tt}(P)= N_{tt} \cdot \frac{\min\limits_{\scriptscriptstyle P_i \in \mathit{Paths}(r)} (\mathit{TravelTime}(P_i))}{\mathit{TravelTime}(P)}
 \label{eq:Eval_tt_2}
 \end{equation}

With this definition, the range of $\mathit{Eval}_{tt}$ is (0,1], being 1 for the path with minimum travel time and decreasing the higher the difference with this. 

\subsubsection{The Evaluation of Congestion}
The behaviour modelled in the agent in this model considers congestion as a negative element for the evaluation of the path. This does not completely reflect the reality, since there could be people who could be attracted by congested paths as well, showing a mere \emph{following} behaviour. On the other hand, by acting on the calibration of the parameter $\kappa_q$ it is possible to define different classes of agents with customized behaviours, also considering attraction to congested paths with the configuration of a negative value. 

For the evaluation of this component of the route decision making activity associated to a path $P$, a function is first introduced for denoting agents $a'$ that precede the evaluating agent $a$ in the route towards the opening $\Omega$ of a path $P$:

\begin{equation}
\begin{aligned}
\mathit{Forward}&(\Omega ,a) =  |\{a' \in Ag\backslash \{a\} : Dest(a') = \Omega\ \land \\
& \mathit{PF}_{\Omega}(\mathit{Pos}(a')) < \mathit{PF}_{\Omega}(\mathit{Pos}(a))\}|
\end{aligned}
\end{equation}

where $Pos$ and $Dest$ indicates respectively the position and current destination of the agent; the fact that $\mathit{PF}_{\Omega}(\mathit{Pos}(a')) < \mathit{PF}_{\Omega}(\mathit{Pos}(a))$ assures that $a'$ is closer to $\Omega$ than $a$, due to the nature of floor fields. Each agent is therefore able to perceive the main direction of the others (its current destination). This kind of perception is plausible considering that only preceding agents are counted, but we want to restrict its application when agents are sufficiently close to the next passage (i.e. they perceive as important the choice of continuing to pursue that path or change it). To introduce a way to calibrate this perception, the following function and an additional parameter $\gamma$ is introduced:

\begin{equation}
\mathit{PerceiveForward}(\Omega ,a) =  \begin{cases}
\mathit{Forward}(\Omega, a), & \mbox{if } \mathit{PF}_{\Omega}(\mathit{Pos}(a))< \gamma \\
0, & \mbox{otherwise}
\end{cases}
\end{equation}

The function $\mathit{Eval}_q$ is finally defined with the normalization of $\mathit{PerceiveForward}$ values for all the openings connecting the region of the agent:

\begin{equation}
\mathit{Eval}_q(P) = N \cdot \frac{\mathit{PerceiveForward}(\mathit{FirstEl}(P), \mathit{myself})}{\mathit{width}(\mathit{FirstEl}(P))}
\end{equation}

where $\mathit{FirstEl}$ returns the first opening to cross of a path, $\mathit{myself}$ denotes the evaluating agent and $\mathit{width}$ scales the evaluation over the width of the door (larger doors sustain higher flows). 

\subsubsection{Propagation of Choices - Following behaviour}
This component of the decision making model aims at representing the effect of an additional stimulus perceived by the agents associated to sudden decision changes of other persons that might have an influence. An additional grid has been introduced to model this kind of event, whose functioning is similar to the one of a dynamic floor field. The grid, called \emph{ChoiceField}, is used to spread a gradient from the positions of agents that, at a given time-step, change their plan due to the perception of congestion.

The functioning of this field is described by two parameters $\rho_c$ and $\tau_c$, which defines the diffusion radius and the time needed by the values to \emph{decay}. The diffusion of values from an agent $a$, choosing a new target $\Omega'$, is performed in the cells $c$ of the grid with $\mathit{Dist}(\mathit{Pos}(a),c) \leq \rho_c$ with the following function:

\begin{equation}
\mathit{Diffuse}(c,a) =
\begin{cases}
1/\mathit{Dist}(\mathit{Pos}(a),c) & \mbox{if } \mathit{Pos}(a) \neq c\\
1 & \mbox{otherwise}
\end{cases}
\end{equation}

The diffused values persist in the \emph{ChoiceField} grid for $\tau_c$ simulation steps, then they are simply discarded. The index of the target $\Omega'$ is stored together with the diffusion values, thus the grid contains in each cell a vector of couples $\{(\Omega_m, \mathit{diff}_{\Omega_m}), \ldots, (\Omega_n, \mathit{diff}_{\Omega_n})\}$, describing the values of influence associated to each opening of the region where the cell is situated. While multiple neighbor agents changes their choices towards the opening $\Omega'$, the values of the diffusion are summed up in the respective $\mathit{diff}_{\Omega'}$. In addition, after having changed its decision, an agent spreads the gradient in the grid for a configurable amount of time steps represented by an additional parameter $\tau_{a}$. In this way it influences the choices of its neighbours for a certain amount of time.

The existence of values $\mathit{diff}_{\Omega_k} > 0$ for some opening $\Omega_k$ implies that the agent is influenced in the evaluation phase by one of these openings, but the probability for which this influence is effective is, after all, regulated by the utility weight $\kappa_f$. In case of having multiple $\mathit{diff}_{\Omega_k} > 0$ in the same cell, a individual influence is chosen with a simple probability function based on the normalized weights $\mathit{diff}$ associated to the cell. Hence, for an evaluation performed by an agent $a$ at time-step $t$, the utility component $Eval_{f}$ can be equal to 1 only for one path $\overline{P}$, between the paths having $\mathit{diff}_{\Omega_k} >0$ in the position of $a$.

\section{Evaluation of the Model with the Entropy Map}
The heat map used to evaluate and calibrate the model of this paper describes the concept of entropy --conceived as in information theory~\cite{BLTJ:BLTJ1338}-- calculated with the set of probabilities of choosing each path from all points of the scenario. Generally speaking, given a set of events $(e_1,e_2, \ldots, e_n)$ and $p$ the function to compute their probability, the entropy of the probability distribution is calculated as:

\begin{equation}
H = \sum\limits_{i=1}^{n} p(e_i) \cdot log_b \frac{1}{p(e_i)}
\end{equation}

where $b$ indicates the measurement unit of the entropy and, as commonly used in this field, we assumed bits ($b=2$). With this definition, $H$ describes the \emph{content of information} of the distribution, but by applying it to the probability distribution of the choices available to an agent $a$ in a certain position during the simulation run, it can provide a description of its \emph{uncertainty} in choosing the path, due to the current dynamics. Applied to our context, the entropy associated to a cell of the environment formally becomes:

\begin{equation}
H(x,y) = \sum\limits_{i=1}^{n} p(P_i) \cdot log \frac{1}{p(P_i)}
\end{equation}

where $\{P_1, \ldots, P_n\}$ is the set of paths from the position $\left( x,y \right)$ and leading to the destination of the agent. For example, let us consider a static setting without the influence of other agents and assume $\kappa_{tt}>0$. While the agent is in a position relatively close to a passage, the probability to choose a path employing that passage will get higher in favour of other possibilities. Hence, the value of $H$ will get closer to 0. Conversely, if the agent is at a relatively same distance between alternative passages, $H$ gets larger values, describing a higher uncertainty of the agent. 

The entropy map is obtained by calculating the entropy of the probability distribution of the choices over possible paths for all the cell of the environment, at a certain step of the simulation. This heat map has already been exploited for the design of the components of this model, as it is explained in~\cite{Crociani_IA2016}. A benchmark test showing the results of the entropy map in different moments of the simulation will be now proposed, with the aim to explain the collective effects of the dynamics on the route choice of agents.

\begin{figure}[t!]
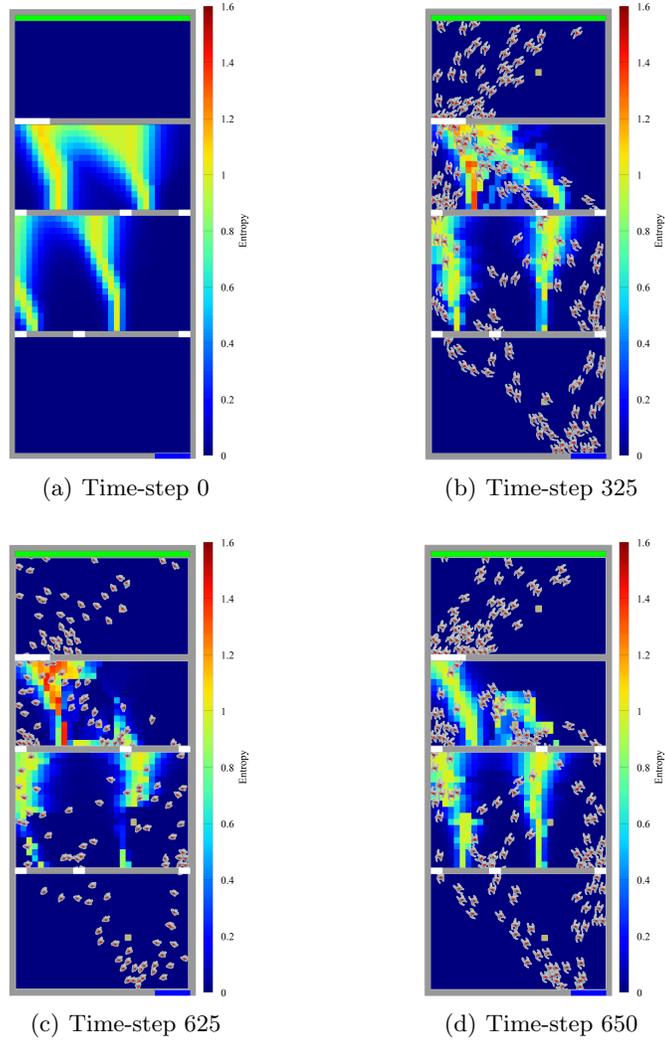

\begin{center}
\subfigure[Time-step 0]{\includegraphics[width=.28\textwidth]{./figs/benchmark_entropies_0_new.png}}\hspace{2cm}
\subfigure[Time-step 325]{\includegraphics[width=.28\textwidth]{./figs/benchmark_entropies_325_new.png}}
\subfigure[Time-step 625]{\includegraphics[width=.28\textwidth]{./figs/benchmark_entropies_625_new.png}}\hspace{2cm}
\subfigure[Time-step 650]{\includegraphics[width=.28\textwidth]{./figs/benchmark_entropies_650_new.png}}
\caption{The entropy maps achieved with the benchmark scenario at different simulation times. Related screenshots of the environment and simulated pedestrians are superimposed to allow the understanding of the dynamics. The objects of the environment are described by the following colors: green for the start area in the top; white for intermediate targets (the bootlenecks connecting the four rooms); grey for obstacles and blue for the destination in the bottom right.}\label{fig:entropies_result}
\end{center}
\end{figure}

\begin{figure}[t!]
\begin{center}
\subfigure[]{\includegraphics[width=.15\textwidth]{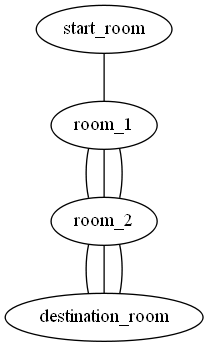}}
\subfigure[]{\includegraphics[width=.8\textwidth]{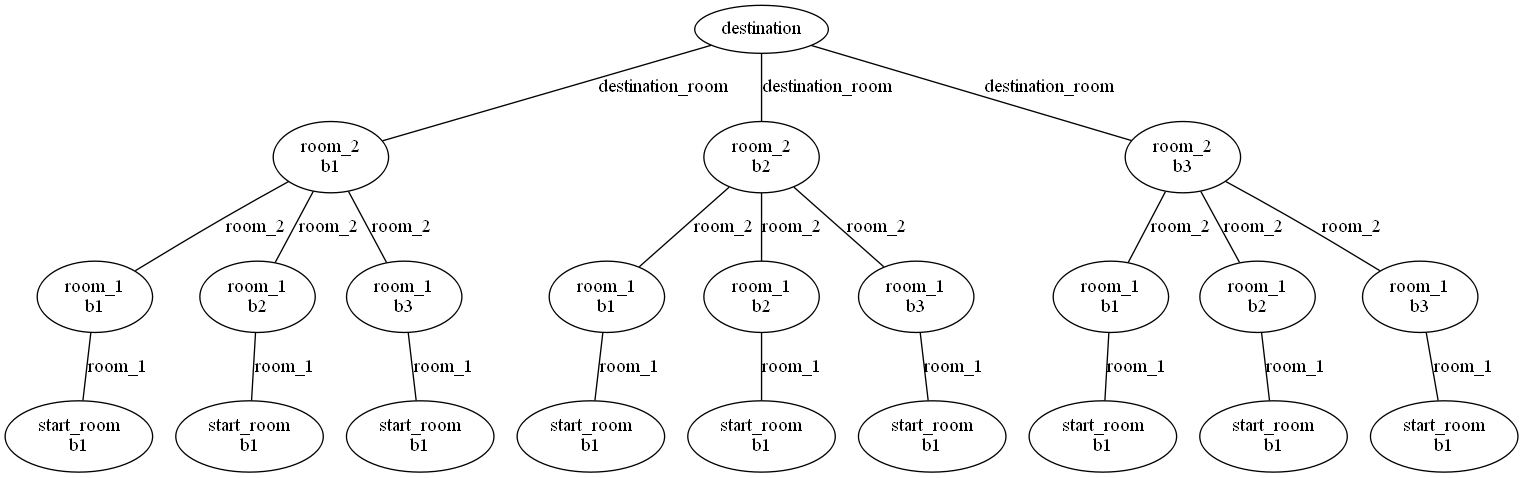}}
\caption{The cognitive map (a) and the paths tree (b) generated for the simulated scenario.}\label{fig:cog_map_paths_tree}
\end{center}
\end{figure}

The environment used for the experiment is shown in Figure~\ref{fig:entropies_result}, together with all entropy maps: the scenario simulates a relatively high incoming flow of pedestrians (an average of 4 persons/s) from the start area, the green object located at the top, and moving towards the destination located at the bottom right of the image. The pedestrian flow crosses four rooms that we denote with numbers describing the order of appearance. The transition between the first and second room is allowed by a single bottleneck of 2.4 m, while the second and third one are managed by three bottlenecks of 0.8 m. To induce more variations in the choice of path during the simulation, the position of bottlenecks describing the second and third transitions is mirrored.

All heat maps proposed in the figure describe the same range of values for the entropy. The colour palette is configured so that dark blue cells indicate a certain choice of a path for the simulated pedestrians located in them. Conversely, a red colour of a cell means that the probability distribution of the available paths is uniform. An arbitrary agent has a maximum of three possible choices in this environment (the three openings providing access to the third and fourth room), thus the maximum theoretical value of entropy is:

$$\sum\limits_{i=1}^{3} p(P_i) \cdot log_2\frac{1}{p(P_i)} = \sum\limits_{i=1}^{3} \frac{1}{3} \cdot log_2(3) = log_2(3) \cong 1.58$$

Figure~\ref{fig:entropies_result}(a) displays the entropy map generated at the very beginning of the simulation. In that time there are still no agents moving in the environment and the spatial distribution of the probability to choose paths is only influenced by the distance towards the destination. This is reflected in the figure, where higher entropy values (i.e. uncertainty of simulated pedestrians) are found in areas describing similar distances for the available paths. Moreover, the asymmetry of the doors location in the environment is also marked in the map. 

Figures~\ref{fig:entropies_result}(b), (c) and (d) show how the entropy distribution is modified by the dynamics and the congestion arisen in front of the bottlenecks. In Figure~\ref{fig:entropies_result}(b) the effect of queueing is particularly marked in the second room and the blue area around the middle passage gets significantly smaller in favour of the righter one. In addition, the values of entropy right after the passage leading to the second room did get higher, meaning that the current situation makes a percentage of agents to choose the righter passage just after entering in the second room: this is very unlikely to happen in a free flow situation, as it can be seen in Figure~\ref{fig:entropies_result}(a).

Finally, Figure~\ref{fig:entropies_result}(c) and (d) exemplify a situation in which some agents switch their decision in favour of a less congested passage. In Figure~\ref{fig:entropies_result}(c) the effects of \emph{ChoiceField} grid is particularly visible in the second and third room: the blue area at the center of the second room is caused by few agents that have changed their decision in favour of the rightest passage, which increases its usage in the successive time-steps (Figure~\ref{fig:entropies_result}(d)). Moreover, a couple of agents in the central-right part of the third room change their route choice in favour of the central passage, leading to a smoother balance of the doors usage in the successive time of the simulation (the ``uncertainty area'' between the central and the rightest passages in the third room is almost a straight line in Figure~\ref{fig:entropies_result}(d), highlighting a well balanced usage of the two doors by the agents). 

\section{Conclusions}
We have shown how the concept of entropy can be used to evaluate uncertainty in probabilistic models that simulate complex dynamics. The above results highlighted that the proposed entropy map is a powerful tool to both support the design of the model and to analyse and understand the effects of the model components in the simulated dynamics. The calibration of the model can also be supported by the usage of the entropy map. On one side, the analysis is able to emphasize too effective mechanisms of the model: for example, the \emph{following behaviour} of this model can trigger a viral effect in the simulated dynamics with high values of the parameter $\kappa_f$. On the other side, components that are believed to be significant for the simulated behaviour should produce low entropy values in the heat map.

Entropy maps can be generalised to other decision modeling situations, especially those in which an agent faces alternative choices related to the spatial structure of the environment in which she is situated. From this perspective, this notion and tool could be relevant, for instance, for agent-based approaches to information foraging~\cite{DBLP:conf/iir/DriasP17} or imitation behaviours in social media.

%\bibliographystyle{splncs04}
%\bibliography{library.bib}

\end{document}